\documentclass[aps,prb,superscriptaddress,preprint,showpacs]{revtex4} 
\usepackage{graphicx} 
\begin{document}  
  
\title{Topological Hunds rules and the electronic properties    
 of a triple lateral quantum dot molecule}  
  
\author{Marek Korkusinski}  
\affiliation{Quantum Theory Group, Institute for Microstructural Sciences,   
        National Research Council of Canada, Ottawa, Ontario, Canada K1A 0R6}

\author{Irene Puerto Gimenez} 
\affiliation{Quantum Theory Group, Institute for Microstructural Sciences,   
        National Research Council of Canada, Ottawa, Ontario, Canada K1A 0R6}

\author{Pawel Hawrylak} 
\affiliation{Quantum Theory Group, Institute for Microstructural Sciences,   
        National Research Council of Canada, Ottawa, Ontario, Canada K1A 0R6}

\author{Louis Gaudreau} 
\affiliation{Quantum Physics Group, Institute for Microstructural Sciences,   
        National Research Council of Canada, Ottawa, Ontario, Canada K1A 0R6}  
\affiliation{R\'egroupement Qu\'eb\'ecois sur les Mat\'eriaux de
  Pointe, Universit\'e de Sherbrooke, Qu\'ebec, Canada J1K 2R1} 

\author{Sergei A. Studenikin} 
\affiliation{Quantum Physics Group, Institute for Microstructural Sciences,   
        National Research Council of Canada, Ottawa, Ontario, Canada K1A 0R6}

\author{Andrew S. Sachrajda }  
\affiliation{Quantum Physics Group, Institute for Microstructural Sciences,   
        National Research Council of Canada, Ottawa, Ontario, Canada K1A 0R6}

\begin{abstract}  
We analyze theoretically and experimentally the electronic structure and 
charging diagram of three coupled lateral quantum dots filled with electrons.   
Using the Hubbard model and real-space exact diagonalization techniques  
we show that the electronic properties of this artificial molecule can be 
understood using a set of topological Hunds rules.  
These rules relate the multi-electron energy levels to  spin  
and the inter-dot tunneling $t$, and control charging energies.  
We map out the charging diagram for up to $N=6$ electrons and predict a  
spin-polarized phase for two holes.  
The theoretical charging diagram is compared  with the measured charging 
diagram of the gated  triple-dot device.  
\end{abstract}  


\maketitle  
  
\section{Introduction}  
  
Following on earlier work which showed that a small and well-controlled number 
of electrons can be confined in a 
single\cite{ciorga_sachrajda_prb2000,tarucha_austing_prl1996} and a double    
quantum 
dot,\cite{holleitner_blick_science2002,pioro_abolfath_prb2005,koppens_folk_science2005,petta_johnson_science2005,hatano_stopa_science2005} 
an artificial lateral quantum molecule consisting of three 
quantum-mechanically coupled lateral quantum dots has been  
demonstrated.\cite{gaudreau_studenikin_prl2006}    
The triple quantum dot molecule is a natural step toward creating 
quantum dot networks, with potential applications in quantum 
computing.\cite{brum_hawrylak_sm1997,loss_divincenzo_pra1998,divincenzo_bacon_nature2000} 
When filled with three electrons, one electron per dot, this device 
can serve as a simple quantum logic circuit, with each electron spin treated 
as a qubit. 
One can also use the molecule as a single coded
qubit,\cite{divincenzo_bacon_nature2000,hawrylak_korkusinski_ssc2005,weinstein_hellberg_pra2005,scarola_park_prl2004} 
whose states are encoded in the states of three electronic spins but
tunable with applied voltage.  
The triple dot could also be used to create entanglement between spin 
qubits,\cite{saraga_loss_prl2003} spin and charge 
qubits,\cite{lehur_recher_prl2006}  as a charge 
rectifier,\cite{stopa_prl2002,vidan_westervelt_apl2004} and may
exhibit a characteristic Kondo effect when coupled to the 
leads.\cite{ingersent_ludwig_prl2005,zitko_bonca_prb2006,zitko_bonca_ramsak_prb2006,jiang_sun_prb2005,kuzmenko_kikoin_prl2006,kuzmenko_kikoin_prb2006,avishai_kuzmenko_physe2005,sakano_kawakami_prb2005} 
With electrons localized on individual dots and their tunneling controlled by 
gates, the triple dot molecule can be also thought of as an implementation 
of the tunable Hubbard model, an important step toward realization of 
``quantum 
materials''.\cite{long_fechrenbacher_jphys1990,elfimov_yunoki_prl2002,anderson_lee_jphys2004,doiron_walker_nature2003,saitoh_okamoto_nature2001,coey_nature2004} 
 
The electronic properties of the triple quantum dot with one electron 
per dot have been studied theoretically by a number of authors. 
To make contact with the pairwise-exchange formalism used in quantum 
information,\cite{divincenzo_bacon_nature2000} attempts were made to 
map the properties of this system onto those of the three-spin 
Heisenberg model.  
Scarola and Das Sarma\cite{scarola_dassarma_pra2005} used the Hubbard, 
variational, and exact diagonalization approaches to demonstrate that 
this mapping can be carried out only for a limited range of 
triple-dot parameters. 
Mizel and 
Lidar\cite{mizel_lidar_prl2004,mizel_lidar_prb2004,woodworth_mizel_jpcm2006} 
arrived at similar conclusions using the Heitler-London and 
Hund-M\"ulliken schemes to calculate the energy levels 
of three coupled dots with one electron per dot. 
In both cases the many-body effects were responsible for the 
appearance of higher-order terms in the effective spin Hamiltonian. 
In an alternative approach, in 
Ref.~\onlinecite{hawrylak_korkusinski_ssc2005} we have used   
real-space wave functions and the configuration-interaction technique 
to analyze the three-electron triple-dot molecule acting as a single 
coded qubit and shown how its energy levels can be tuned by voltages 
applied to gates defining the structure. 
 
Properties of the triple-dot molecule as a scattering center  
have also been studied using quantum transport techniques. 
Using the density-functional and quantum Monte Carlo methods, 
Stopa\cite{stopa_prl2002} calculated the current flowing through a 
nominally empty molecule connected to electron reservoirs and under 
bias.  
The rectifying behavior of the system predicted in this analysis was 
confirmed experimentally.\cite{vidan_westervelt_apl2004} 
Landr\'on de Guevara and Orellana\cite{guevara_orellana_prb2006} 
calculated the zero-temperature conductance through a linear molecule 
coupled in parallel to the leads using a Hubbard approach in a 
magnetic field.  
Apart from the Fano resonances in the spectrum, they found evidence of 
formation of the quantum-molecular states decoupled from the leads. 
The Hubbard model has also been used to investigate the triple-dot 
system in the Kondo regime, both in the 
linear\cite{zitko_bonca_prb2006,zitko_bonca_ramsak_prb2006,jiang_sun_prb2005} 
and triangular 
topology.\cite{kuzmenko_kikoin_prl2006,kuzmenko_kikoin_prb2006,avishai_kuzmenko_physe2005,sakano_kawakami_prb2005} 
 
In this paper we describe the electronic properties of a lateral  
triple quantum dot molecule as a function of electron numbers. 
In analogy to the work on quantum 
materials,\cite{long_fechrenbacher_jphys1990,elfimov_yunoki_prl2002}  
we model our system with the Hubbard Hamiltonian, but the obtained
results are verified by microscopic methods.  
In the Hubbard model  we retain only one lowest-energy orbital per
dot.  
The lowest-energy shell of the molecule can be filled with up to
$N_e=6$ electrons.   
We analyze in detail the ordering of energy levels, the spacing of 
Coulomb blockade peaks and the charging and spin phase diagram of this
shell.    
We demonstrate that the energy levels of the molecule are related to 
the total spin of electrons but not directly related to the charge $e$.   
We find the spin singlet as the two-electron ground state, with the 
singlet-triplet (S-T) splitting proportional to the single-particle 
tunneling matrix element $t$.   
This is in contrast to atoms, where the S-T splitting is proportional   
to the electronic exchange and hence to $e^2$, or to magnetic solids, 
where super-exchange leads to S-T splitting proportional to $1/e^2$.  
On the other hand, for two holes ($N_e=4$) we predict a spin 
polarized ground state and a singlet-triplet transition driven only by 
modifying the topology of the system.   
For three electrons in a half-filled shell ($N_e=3$) we confirm the 
existence  of the frustrated antiferromagnetic ground 
state.\cite{hawrylak_korkusinski_ssc2005}   
The fact that the tunneling alone distinguishes singlet and triplet 
states is related to the interplay of the Fermi statistics and system  
topology.  
We term the set of rules established here and relating spin of the
ground state to the filling of the shell, topology, and tunneling,
"topological Hunds rules".  
The ability to tune tunneling by gates opens the 
possibility of directly manipulating the electron spin using 
electrical means only, of interest in designing novel quantum 
materials, magneto-electronics and quantum computation.   
We show that the Hubbard model is capable of reproducing the charging   
diagram of a lateral gated triple-dot measured recently by Gaudreau  
{\em et al.}\cite{gaudreau_studenikin_prl2006}   
 
The paper is organized as follows.   
In Sec.~\ref{secmodel} we describe the model lateral triple-dot device  
and construct the Hubbard Hamiltonian.  
In Sec.~\ref{sec1to6} we determine the electronic  
structure of the device charged with $N_e=1$ to $6$ electrons.   
Results of the Hubbard model are tested against real space (RSP) 
configuration interaction (RSP-CI) and linear combination of atomic or 
quantum dot orbitals (LCAO-CI) calculations.   
The charging diagram as a function of the  dot energies is  
presented and analyzed in Sec.~\ref{seccharging}.  
In Sec.~\ref{secexperiment} we relate the calculated  
and measured charging diagrams.  
Summary and conclusions are presented in Sec.~\ref{summary}. 
  
\section{The model \label{secmodel}}  
  
The proposed model gated triple-dot device realizing the triple dot 
using only metallic gates, studied in 
Ref.~\onlinecite{hawrylak_korkusinski_ssc2005} and related to the one 
studied by Gaudreau {\em et al.} in 
Ref.~\onlinecite{gaudreau_studenikin_prl2006}, is shown in  
Fig.~\ref{fig1}(a).      
It consists of a heterojunction with a two-dimensional electron gas 
(2DEG) created at a distance $D$ below  the top surface of the sample.    
The metallic gates deposited on the surface serve to deplete the 2DEG  
underneath.  
Any opening in the gates is translated electrostatically into a local  
potential minimum, capable of confining a small number of electrons.  
Thus, in our model the three circular holes in the main gate (shown  
in gray) define a triangular triple quantum dot lateral confinement.  
Each isolated potential minimum gives rise to a quantized energy 
spectrum, of which we retain only the lowest energy level $E_i$ in dot 
$i$. 
By tuning the voltage on the main gate we can control the number of  
confined electrons.  
For example, in Fig.~\ref{fig1}(a) we show $N_e=2$ electrons with  
parallel spins localized on two of the dots.  
This is not, however, a depiction of a quantum molecular state:  
due to the interdot coupling the electrons are delocalized across the  
molecule.  
The main gate alone defines a symmetric triangular molecule with  
identical pairwise coupling of all dots.  
This triple-dot potential can be well approximated by a sum of three 
Gaussians.  
 
The single-particle confinement can be additionally tuned by three  
smaller gates, shown in red, green, and blue.    
Their arrangement with respect to the potential minima is shown  
schematically in Fig.~\ref{fig1}(b).   
The gate $VG_1$ controls simultaneously the lowest energy levels $E_1$  
and $E_2$ of dots 1 and 2, and the gate $VG_3$ controls the energy  
level $E_3$ of dot 3.  
Additionally, the gate $VG_{13}$ is designed to tune the topology  
of the system without significantly changing the energies $E_i$.  
By biasing it with a sufficiently high negative voltage we increase  
the tunneling barrier between dots 1 and 3 and change the sample  
layout from a closed triangle, in which all dots are identically coupled,  
to a linear molecule, in which the tunneling between dots 1 and 3 is  
not allowed.  
  
We examine the electronic properties of our triple quantum dot  
molecule in the frame of the Hubbard model with one spin-degenerate  
orbital per dot. Without specifying them explicitly, the localized 
orbitals in the Hubbard model are assumed to be orthogonal.  
This is to be contrasted with the approach starting from the linear 
combination of atomic orbitals (LCAO), which are non-orthogonal.   
The orthogonalization leads to extended, quantum-molecular  orbitals 
which serve as a basis for CI calculation.     
In the Hubbard model, with  $c_{i\sigma}^+$ ($c_{i\sigma}$) operators 
creating (annihilating) electrons with spin $\sigma$ on the orbital of 
$i$-th dot, the Hamiltonian can be written as:   
\begin{equation}  
  \hat{H} =   
  \sum\limits_{\sigma,i=1}^{3} E_i c_{i\sigma}^+ c_{i\sigma}  
+ \sum\limits_{\sigma,i,j=1,i\neq j}^{3}   
             t_{ij} c_{i\sigma}^+ c_{j\sigma}   
+ \sum\limits_{i=1}^{3} U_i n_{i\downarrow} n_{i\uparrow}  
+ {1\over{2}} \sum\limits_{i,j=1,i\neq j}^{3} V_{ij} \varrho_i \varrho_j,  
\label{hubhamil}  
\end{equation}  
where  $n_{i\sigma} = c_{i\sigma}^+ c_{i\sigma}$   
and    
$\varrho_i = n_{i\downarrow} + n_{i\uparrow}$  
are, respectively, the spin and charge density on the $i$-th dot.   
The above Hamiltonian is characterized  by   
the energy levels of the $i$-th quantum dot $E_i$,   
the tunneling matrix elements $t_{ij}$ between dots $i$ and $j$,    
the on-site Hubbard repulsion $U_i$, and the direct Coulomb matrix  
elements $V_{ij}$ between dots $i$ and $j$.  
These Hubbard parameters are schematically shown in  
Fig.~\ref{fig1}(b).   
With one energy level per dot the triple-dot molecule can be filled  
with up to $N_e=6$ electrons.

\section{Electronic structure of the triple dot with 1 to 6 electrons  
  \label{sec1to6}}   
  
\subsection{One electron and one hole}  
  
We look for the eigenenergies and eigenvectors of the Hamiltonian  
(\ref{hubhamil}) using the exact diagonalization approach.  
To this end, we create all possible configurations of $N_e$ electrons  
on the three localized orbitals, write the Hamiltonian matrix in this  
basis, and diagonalize it numerically.  
In the simplest case of $N_e=1$ the basis contains three  
non-overlapping states, $\{|1\rangle,|2\rangle,|3\rangle\}$, where   
$|i\rangle = c^+_{i\downarrow}|0\rangle$ and $|0\rangle$ denotes the  
vacuum.  
In this basis the diagonal Hamiltonian matrix elements are $\langle  
i|\hat{H}|i\rangle = E_i$ and the off-diagonal elements $\langle  
i|\hat{H}|j\rangle=t_{ij}$.   
With the three dots on resonance, i.e., with $E_1=E_2=E_3=E$ and  
$t_{12}=t_{23}=t_{13}=t$, the one-electron energy spectrum is composed  
of one level with energy $E_A=E+2t$, and one doubly-degenerate level with  
energy $E_B=E_C=E-t$.   
The order of these levels depends on the sign of the element $t$.  
In numerical calculations of the single-particle spectrum  
corresponding to the potential produced by metallic  
gates\cite{hawrylak_korkusinski_ssc2005}  shown in Fig.~\ref{fig1}(a)   
we find the ground state to be non-degenerate, indicating that $t<0$.   
Additionally, the magnitude of the tunneling matrix element can be  
found from the single-particle energy gap $\Delta =3|t|$.  
  
Knowledge of the sign of the off-diagonal element allows us to  
construct the single-particle molecular orbitals.  
The ground state is $|M_1\rangle = {1\over\sqrt{3}}\left(|1\rangle +  
  |2\rangle + |3\rangle  \right)$, while the two degenerate excited  
states are  
$|M_2\rangle = {1\over\sqrt{2}}\left( |1\rangle - |2\rangle \right)$  
and  
$|M_3\rangle = 
{1\over\sqrt{6}}\left(|1\rangle+|2\rangle-2\cdot|3\rangle\right)$.  
The states $|M_2\rangle$ and $|M_3\rangle$ were chosen to be symmetric  
with respect to a mirror plane passing through the dot 3 and  
intersecting the $(1-2)$ base of the triangle at its midpoint.  
However, due to the degeneracy of the two levels, any pair of  
orthogonal states created as linear combinations of $|M_2\rangle$ and  
$|M_3\rangle$ will be viable as eigenstates.   
The degeneracy of the excited states is a direct consequence of the  
symmetry of the triangular molecule.  
Changing its topology, e.g., by increasing the tunneling barrier  
between dots 1 and 3, will remove the degeneracy.  
In the limit of an infinite barrier, i.e., $t_{13}=0$, we deal with a  
linear triple-dot molecule, whose single-particle energy spectrum  
consists of three equally spaced levels:  
$(E-\sqrt{2}|t|,E,E+\sqrt{2}|t|)$.   
Thus, the triangular triple dot design makes it possible to engineer  
the degeneracy of states solely by electrostatic means.  
  
Now we can start to populate our triple-dot molecule with electrons.  
Let us start our many-body analysis with the simplest case of  
$N_e=5$.   
As the maximal number of electrons in our system is six, we can  
interpret the five-electron configurations as those of a single hole.  
The hole (e.g., with spin down) can be placed on either of the dots,  
and thus our basis consists of three configurations:  
$|1^{(H)}\rangle=h^+_{1\downarrow}|N_e=6\rangle =  
c^+_{3\uparrow}c^+_{2\uparrow}c^+_{3\downarrow}c^+_{2\downarrow}  
c^+_{1\downarrow}|0\rangle$,  
$|2^{(H)}\rangle=h^+_{2\downarrow}|N_e=6\rangle =  
c^+_{1\uparrow}c^+_{3\uparrow}c^+_{3\downarrow}c^+_{2\downarrow}  
c^+_{1\downarrow}|0\rangle$,  
and  
$|3^{(H)}\rangle=h^+_{3\downarrow}|N_e=6\rangle =  
c^+_{2\uparrow}c^+_{1\uparrow}c^+_{3\downarrow}c^+_{2\downarrow}  
c^+_{1\downarrow}|0\rangle$,  
with $h^+_{i\sigma}$ being the creation operator of the hole with spin  
$\sigma$ on the $i$-th dot.  
It is convenient to express the energies of these configurations with  
respect to the total energy of the system with six electrons  
$E_F=2E_1+2E_2+2E_3 + U_1+U_2+U_3 + 4V_{12}+4V_{13}+4V_{23}$.  
We have then  
$E_1^{(H)}=E_F-E_1-U_1-2V_{12}-2V_{13}$,  
$E_2^{(H)}=E_F-E_2-U_2-2V_{12}-2V_{23}$, and  
$E_3^{(H)}=E_F-E_3-U_3-2V_{13}-2V_{23}$.  
The three energies are respectively the diagonal terms of our  
single-hole Hamiltonian.  
The off-diagonal terms are composed out of the  
single-particle tunneling matrix elements.  
We have $\langle i^{(H)}|\hat{H}|j^{(H)}\rangle = -t_{ij}$; the  
negative phase is due to the anticommutation relations of the  
electronic creation and annihilation operators.  
As we can see, the single-hole Hamiltonian can be obtained from the  
single-electron Hamiltonian by appropriately modifying the diagonal  
terms and setting $t_{ij}\leftrightarrow -t_{ij}$.  
This is the signature of the particle-hole
symmetry.\cite{long_fechrenbacher_jphys1990}   
However, for the triangular triple dot on resonance this symmetry is  
not reflected in the energy spectrum of the hole:  
in this case, the opposite sign of the off-diagonal element leads to a   
doubly-degenerate hole ground state.  
This property is immediately apparent in the molecular basis: we  
create the lowest-energy configuration by filling the molecular ground  
state $|M_1\rangle$ with two of the five electrons, and distributing the  
remaining three on the degenerate orbitals $|M_2\rangle$ and  
$|M_3\rangle$.   
The latter can be accomplished in two energetically equivalent ways,  
hence the double degeneracy.  
Note, however, that the electron-hole symmetry is fully restored upon  
transition to the linear triple-dot molecule.  
For this topology, the single-particle spectrum of both the electron  
and the hole consists of three equally spaced non-degenerate levels.

\subsection{Two electrons and two holes}  
  
The interplay of topology and statistics is particularly important in  
the cases of two electrons and two holes confined in the triple dot  
molecule.   
Let us consider the case of $N_e=2$ first.  
Since the Hamiltonian (\ref{hubhamil}) commutes with the total spin  
operator, we can classify the two-electron states into singlets and  
triplets.  
Working with the molecular basis set, we form the configuration with  
the lowest energy by placing both carriers with antiparallel spins on  
orbital $|M_1\rangle$.   
Therefore we expect the ground state of the two-electron system to be  
a spin singlet, irrespective of the molecule's topology.  
However, in order to examine the topological and statistical effects  
in the energy spectrum and the structure of the wave functions, we  
carry out a systematic analysis in the localized basis.  
  
Due to Fermi statistics, the two electrons with parallel spins cannot   
occupy the same quantum dot.  
Hence there are only three possible triplet configurations,  
$|T_1\rangle=c^+_{2\downarrow}c^+_{1\downarrow}|0\rangle$,  
$|T_2\rangle=c^+_{3\downarrow}c^+_{1\downarrow}|0\rangle$,  
and  
$|T_3\rangle=c^+_{3\downarrow}c^+_{2\downarrow}|0\rangle$,  
shown schematically in Fig.~\ref{fig2}(a).   
The three triplet configurations interact with each other only via the  
single-particle tunneling Hamiltonian.  
However, in evaluating the respective matrix elements we need to  
follow the Fermionic anticommutation rules of the creation and  
annihilation operators. 
For example, acting with $\hat{H}$ on the configuration $T_1$ to  
produce the configuration $T_3$ requires the evaluation of the  
following expression:    
$\hat{H}|T_1\rangle=+t_{31}c^+_{3\downarrow}c_{1\downarrow}  
c^+_{2\downarrow}c^+_{1\downarrow}|0\rangle$.  
In order to remove the electron 1 we first have to move it around  
electron 2, and so $\hat{H}|T_1\rangle  
=-t_{31}c^+_{3\downarrow}c^+_{2\downarrow}  
c_{1\downarrow}c^+_{1\downarrow}|0\rangle  
= -t_{31}|T_3\rangle$.  
Hence, tunneling of the electron from dot $1$ to dot $3$ in the presence  
of the electron in dot $2$ generates an additional phase or changes 
the sign of the tunneling matrix element.  
This is of course the most elementary property of Fermions brought out   
so clearly in this simple model.  
By contrast, tunneling from dot $2$ to dot $3$ in the presence of electron   
in dot $1$ does not change   
the sign of the tunneling matrix element.  
The resulting triplet Hamiltonian matrix takes the following form:  
\begin{equation}  
\hat{H}_T = \left[  
  \begin{array}{ccc}  
    E_1+E_2+V_{12} & t_{23} & -t_{13}\\  
    t_{23} & E_1+E_3+V_{13} & t_{12} \\  
    -t_{13} & t_{12} & E_2+E_3+V_{23}\\  
  \end{array}  
\right].  
\label{triphamil}  
\end{equation}  
$\hat{H}_T$ is related to the one-hole Hamiltonian.  
This similarity becomes more apparent if $\hat{H}_T$ is written in the  
basis $\{ |T_1\rangle, -|T_2\rangle, |T_3\rangle  \}$, in which case  
all the off-diagonal elements acquire a negative phase.  
This is not surprising, since the single-hole configurations analyzed  
in the previous Section can be generated from the above triplet  
configurations simply by adding to them an inert core of three  
electrons spin up, one electron per dot.  
With the three dots on resonance and all tunneling matrix elements  
$t_{ij}$ equal and negative, the triplet energy   
spectrum is found to be $(2E+V-|t|,2E+V-|t|,2E+V+2|t|)$.   
As in the case of the single hole, the lowest-energy triplet state is  
doubly degenerate.  
Moreover, the renormalization of the lowest energy $2E+V-|t|$ from the  
single configuration energy $2E+V$, as well as the gap in the triplet  
spectrum, are determined entirely by tunneling.    
The splitting between the ground and first excited states is the same  
as that found in the single-carrier case and equals $3|t|$.  
  
We shall now demonstrate that topology and statistics  
differentiates between triplet and singlet two-electron states.   
The singly-occupied singlet configurations $|S_1\rangle$,  
$|S_2\rangle$, and $|S_3\rangle$ are obtained from the triplet  
configurations $|T_1\rangle$, $|T_2\rangle$, and $|T_3\rangle$ by  
flipping the spin of one electron and properly antisymmetrizing the  
configurations.   
For example, the configuration  
$|S_1\rangle={1\over\sqrt{2}}\left(c^+_{2\downarrow}c^+_{1\uparrow}+  
 c^+_{1\downarrow}c^+_{2\uparrow}  \right)|0\rangle$.   
In addition to the singly-occupied configurations there are also three  
doubly-occupied configurations, e.g.,  
$|S_4\rangle=c^+_{1\downarrow}c^+_{1\uparrow}|0\rangle$,   
as shown in Fig.~\ref{fig2}(b).   
In the basis of the six configurations the two-electron singlet  
Hamiltonian can be written as:  
\begin{equation}  
\hat{H}_S = \left[  
  \begin{array}{cccccc}  
    E_1+E_2+V_{12} & t_{23} & t_{13} & \sqrt{2}t_{12}& \sqrt{2}t_{12}  
    & 0\\  
    t_{23} & E_1+E_3+V_{13} & t_{12} & \sqrt{2}t_{13} & 0 &  
    \sqrt{2}t_{13}\\   
    t_{13} & t_{12} & E_2+E_3+V_{23} & 0 & \sqrt{2}t_{23} &  
    \sqrt{2}t_{23}\\   
    \sqrt{2}t_{12}& \sqrt{2}t_{13} & 0 & 2E_1 + U_1 & 0 & 0 \\  
    \sqrt{2}t_{12}& 0 & \sqrt{2}t_{23} & 0 & 2E_2 + U_2 & 0 \\  
    0 & \sqrt{2}t_{13} & \sqrt{2}t_{23} & 0 & 0 & 2E_3 + U_3 \\  
  \end{array}  
\right].  
\label{singhamil}  
\end{equation}  
The $3\times3$ upper left-hand corner of $\hat{H}_S$  
corresponds to the three singly occupied configurations $|S_1\rangle$,  
$|S_2\rangle$, and $|S_3\rangle$.    
It is similar to the two-electron triplet Hamiltonian $\hat{H}_T$  
but differs from it by the positive phase of the tunneling matrix  
element $t_{13}$.    
Hence, in the triangular topology of the triple-dot molecule the  
tunneling from dot 1 to dot 3 distinguishes between the singlet and  
the triplet spin configurations.    
By setting $t_{13}=0$, i.e., upon transition to the linear topology,  
this difference disappears.  
However, the singlet basis is still different from its triplet  
counterpart due to the presence of the doubly-occupied configurations.  
  
For the dots on resonance the energies of the six singlet levels can  
be obtained analytically.  
The spectrum can be grouped into two non-degenerate levels $E^S_{1,2}$:  
\begin{equation}  
  E^S_{1,2} = (2E+V-2|t|)   
  + {1\over2} \left[  
    (U-V+2|t|)\pm\sqrt{(4\sqrt{2}t)^2+(U-V+2|t|)^2}  
  \right],  
  \label{enesinglow}  
\end{equation}  
and two groups of doubly degenerate levels $ E^S_{3-6}$:  
\begin{equation}  
  E^S_{3-6} = (2E+V+|t|)   
  + {1\over2} \left[  
    (U-V-|t|)\pm\sqrt{(2\sqrt{2}t)^2+(U-V-|t|)^2}  
  \right].  
  \label{enesinghigh}  
\end{equation}  
In the strong coupling limit $U\gg V > |t|$ the singlet ground-state  
energy $E^S_1\approx (2E+V-2|t|)-{8t^2\over U-V}$, while the triplet  
energy $E^T_1 = (2E+V-|t|)$.  
Thus, the two-electron ground state is always a spin singlet.  
The singlet-triplet gap, separating $E^S_1$ from $E^T_1$ is  
$\Delta^{S-T} \approx |t| + {8t^2\over U-V}$.  
It is proportional to the tunneling matrix element $|t|$ and contains  
the second-order super-exchange correction $\sim t^2/(U-V)$ due to the  
doubly occupied singlet configurations.  
Removing the resonance by detuning the onsite energies $E_i$ enhances  
the contribution from the doubly-occupied states.  
Therefore the ground state maintains its singlet character  
independently of the choice of gate voltages.  
  
The situation is qualitatively different when two holes,   
instead of two electrons, populate the system.   
The two holes are created when two electrons are removed from the  
closed-shell configuration with $N_e=6$, i.e., they correspond to  
$N_e=4$ electrons.   
In the molecular basis corresponding to the triangular triple dot we  
put two electrons on the lowest-energy orbital $|M_1\rangle$, and the  
remaining two electrons on the degenerate pair of orbitals  
$|M_2\rangle$ and $|M_3\rangle$.   
With this alignment of levels it is possible to create both triplet  
and singlet configurations, all with the same single-particle energy,  
and it is not immediately clear which total spin is preferred.  
On the other hand, in the limit of the linear triple dot the molecular  
orbitals are non-degenerate and the four-electron ground state is  
expected to be a spin singlet.  
  
The selected two-hole singlet and triplet configurations in the localized  
basis are illustrated in Fig.~\ref{fig2}(c).   
Let us focus on the triplets first.  
They involve one electron spin-up occupying the first, second, or  
third dot in the presence of an inert core of three spin-down electrons.  
For example, the configuration shown in left-hand panel of  
Fig.~\ref{fig2}(c) can be written as   
$|T^{(H)}_1\rangle = h^+_{1\downarrow}h^+_{2\downarrow}|N_e=6\rangle  
=c^+_{3\uparrow}c^+_{3\downarrow}c^+_{2\downarrow}c^+_{1\downarrow}|0\rangle$.  
Therefore, the hole triplet Hamiltonian is equivalent to the  
single-electron Hamiltonian, differing from it only in diagonal terms.  
For example, the energy of the configuration $|T^{(H)}_1\rangle$ is   
$ \langle T^{(H)}_1|\hat{H}|T^{(H)}_1\rangle =   
E_F - E_1 - E_2 - U_1 - U_2 - 3V_{12} - 2V_{13} - 2V_{23}$.  
The two-hole triplet Hamiltonian can also be compared to the   
two-electron triplet Hamiltonian $\hat{H}_T$, written  
in the modified basis set $\{ |T_1\rangle, -|T_2\rangle, |T_3\rangle  
\}$ (i.e., with all off-diagonal matrix elements acquiring a negative  
phase).  
Setting aside the diagonal matrix elements, the two Hamiltonians are  
connected by the electron-hole symmetry transition  
$t_{ij}\leftrightarrow -t_{ij}$.   
However, unlike that of the electronic triplet, the ground state of  
the hole triplet is non-degenerate, and its energy is  
$E^{T(H)}_1 =  E_F-2E-2U-7V - 2|t|$.   
As it is in the case of the single electron and the single hole,  
the particle-hole symmetry between the two-electron triplet and the  
two-hole triplet is fully restored upon transition to the linear  
topology of the triple dot.  
  
Let us move on to considering the two-hole singlet configurations.  
The singly-occupied states, illustrated in the middle panel of   
Fig.~\ref{fig2}(c), involve the two holes occupying two different  
dots, while the doubly-occupied states, such as the one in the  
right-hand panel of Fig.~\ref{fig2}(c), hold both holes on the same  
dot.  
The two-hole singlet Hamiltonian is analogous to that of the  
two-electron singlet, Eq.~(\ref{singhamil}).   
However, we need to replace the energy of two-electron complexes with  
the energy of two-hole complexes, and change the phase of the  
off-diagonal elements connecting the singly-occupied configurations.   
The sign of elements $\sqrt{2}t_{ij}$ connecting the singly- and  
doubly-occupied configurations does not change, which breaks the  
particle-hole symmetry.  
  
The ground-state energy of the hole singlet for the triangular triple  
dot on resonance is well approximated by  
$E^{S(H)}_1\approx (E_F-2E-2U-7V-|t|)-{2t^2\over U-V}$.   
Compared to the energy of the triplet, $E^{S(H)}_1$ is increased by  
the tunneling element $|t|$, but decreased by the super-exchange  
contribution ${2t^2\over U-V}$.  
Note that the two-hole super-exchange term is four times smaller than  
the super-exchange correction to the energy of the two-electron  
singlet.   
By increasing the tunneling or decreasing the on-site Hubbard repulsion  
we can increase the contribution from super-exchange and lower the energy  
of the singlet state.   
Therefore, the total spin of the two-hole ground state for the triple  
dot on resonance depends on the interplay of Hubbard parameters.  
For $2|t|<U-V$ the ground state is a spin triplet, and a  
triplet-singlet transition can be induced by increasing the hopping   
matrix element.    
The triplet-singlet transition can also be induced by biasing one of  
the dots, which lowers the energy of the doubly-occupied singlet  
configurations.   
Hence the configuration of two holes shows a nontrivial dependence on  
tunneling, Coulomb interactions and gate voltages allowing to control  
the system's magnetic moment purely by electrical means.  
  
\subsection{Three electrons}  
  
To complete our understanding of the energy levels of a triple quantum  
dot molecule we need to analyze the half-filled case of three  
electrons (or, equivalently, three holes).   
We start with the completely spin-polarized system, i.e., one with  
total spin $S=3/2$.   
In this case we can distribute the electrons on the three dots in only  
one way: one electron on each site, which gives a spin-polarized state   
$|a_{3/2}\rangle=c^+_{3\downarrow}c^+_{2\downarrow}c^+_{1\downarrow}|0\rangle$.   
As the basis of our Hilbert space consists of one configuration only,  
$|a_{3/2}\rangle$ is the eigenstate of our system, and its energy is   
$E_{3/2}=E_1+E_2+E_3+V_{12}+V_{13}+V_{23}$.   
Let us now flip the spin of one of the electrons.   
This electron can be placed on any orbital, and with each specific 
placement the remaining two spin-down electrons can be distributed in 
three ways. 
For example, Fig.~\ref{fig3}(a) shows the three configurations with 
the spin-up electron occupying the dot 1. 
Thus, altogether we can generate nine different configurations.   
Three of these configurations involve single occupancy of the  
orbitals.   
They can be written as   
$|a\rangle=c^+_{3\downarrow}c^+_{2\downarrow}c^+_{1\uparrow}|0\rangle$,  
$|b\rangle=c^+_{1\downarrow}c^+_{3\downarrow}c^+_{2\uparrow}|0\rangle$,  
and  
$|c\rangle=c^+_{2\downarrow}c^+_{1\downarrow}c^+_{3\uparrow}|0\rangle$.  
Out of these three configurations we construct the three eigenstates  
of the total spin operator.   
One of those eigenstates is   
$|a_{3/2}\rangle={1\over\sqrt{3}}(|a\rangle + |b\rangle + |c\rangle)$,  
and it corresponds to the total spin $S=3/2$.   
The two other eigenstates,    
$|a_{1/2}\rangle={1\over\sqrt{2}}(|a\rangle - |b\rangle)$  
and  
$|b_{1/2}\rangle={1\over\sqrt{6}}(|a\rangle + |b\rangle - 2|c\rangle)$,  
correspond to the total spin $S = 1/2$.   
The remaining six configurations involve doubly-occupied  
orbitals.   
They are   
$|c_{1/2}\rangle=c^+_{2\downarrow}c^+_{1\downarrow}c^+_{1\uparrow}|0\rangle$,  
$|d_{1/2}\rangle=c^+_{3\downarrow}c^+_{1\downarrow}c^+_{1\uparrow}|0\rangle$,  
$|e_{1/2}\rangle=c^+_{3\downarrow}c^+_{2\downarrow}c^+_{2\uparrow}|0\rangle$,  
$|f_{1/2}\rangle=c^+_{1\downarrow}c^+_{2\downarrow}c^+_{2\uparrow}|0\rangle$,  
$|g_{1/2}\rangle=c^+_{1\downarrow}c^+_{3\downarrow}c^+_{3\uparrow}|0\rangle$,  
$|h_{1/2}\rangle=c^+_{2\downarrow}c^+_{3\downarrow}c^+_{3\uparrow}|0\rangle$.  
All these configurations are eigenstates of the total spin with  
$S=1/2$.    
Thus, among our nine spin-unpolarized states we have one high-spin,  
and eight low-spin states.   
In this basis the Hamiltonian matrix is block-diagonal, with the  
high-spin state completely decoupled.   
The energy corresponding to this state is equal to that of the fully  
polarized system discussed above, and is equal to $E_{3/2}$.   
In the basis of the nine $S=1/2$ configurations we construct the  
Hamiltonian matrix by dividing $9$ configurations into three groups,  
each containing one of the singly-occupied configurations $|a\rangle$,  
$|b\rangle$, and $|c\rangle$.   
By labeling each group with the index of the spin-up electron, the  
Hamiltonian takes the form of a $3\times 3$ matrix:  
\begin{equation}  
  \hat{H}_{1/2}=\left[  
    \begin{array}{ccc}  
      \hat{H}_1 & \hat{T}_{12} & \hat{T}_{31}^+\\  
      \hat{T}_{12}^+ & \hat{H}_2 & \hat{T}_{23}\\  
      \hat{T}_{31} & \hat{T}_{23}^+ & \hat{H}_3\\  
    \end{array}  
    \right].  
    \label{hubbard3e}  
\end{equation}  
The diagonal matrix, e.g.,  
$$  
\hat{H}_1 = \left[  
  \begin{array}{ccc}  
    2E_1+E_2+2V_{12}+U_1 & t_{23} & -t_{13}\\  
    t_{23} & 2E_1+E_3+2V_{13}+U_1 & t_{12} \\  
    -t_{13} & t_{12} & E_1+E_2+E_3+V_{12}+V_{13}+V_{23} \\  
  \end{array}  
\right]  
$$  
describes the interaction of three configurations which contain   
spin-up electron on site 1, i.e., two doubly-occupied configurations  
$|c_{1/2}\rangle$ and $|d_{1/2}\rangle$, and a singly-occupied  
configuration  $|a\rangle$ (in this order, see Fig.~\ref{fig3}(a)).    
The configurations with double occupancy acquire the diagonal  
interaction term $U$.   
The three configurations involve a pair of spin-polarized electrons  
(spin triplet) moving on a triangular plaquette in the presence of a   
``spectator'' spin-up electron.   
Because of the triplet character of the two electrons, the phase of  
the hopping matrix element $-t_{13}$ from site 1 to site 3 is  
different from the phase of the hopping matrix element $+t_{23}$  from  
site 2 to site 3.    
As discussed above, the negative phase in $-t_{13}$ distinguishes  
the singlet and triplet electron pairs.    
The remaining matrices corresponding to spin-up electrons localized on  
sites 2 and 3 can be constructed in a similar fashion.   
The interaction between them is given in terms of effective hopping  
matrix   
$$  
\hat{T}_{ij} = \left[  
  \begin{array}{ccc}  
    0 & -t_{ij} & 0\\  
    0 & 0 & -t_{ij}\\  
    +t_{ij} & 0 & 0\\  
  \end{array}  
\right].  
$$   
There is no direct interaction between the configurations with single  
occupancy, since such scattering process would have to involve two  
electrons, one with spin up and one with spin down.   
This cannot be accomplished by the single-particle tunneling.   
These states are coupled only indirectly, involving the configurations  
with double occupancy.    
  
The low-energy spectrum of the Hubbard Hamiltonian of the $N_e=3$  
quantum-dot molecule can be further approximated by the spectrum of  
the model spin Hamiltonian :  
\begin{equation}  
  H_{3e}=E_{3/2}+ \sum\limits_{i<j}J_{ij}\left(\vec{S}_i\cdot\vec{S}_j  
    - 1/4 \right)+\sum\limits_{i<j<k} D_{ijk} \vec{S}_i\cdot   
\left(  \vec{S}_j \times \vec{S}_k \right) 
\label{heishamil}
\end{equation}  
Here, $E_{3/2}$ is the energy of the spin $S=3/2$ state, $J_{ij}$  
are exchange matrix elements of the Heisenberg part of the spin 
Hamiltonian which depend on microscopic parameters of the triple dot, 
and $D_{ijk}$ are higher order spin-spin interactions discussed, 
e.g., by Scarola and Das Sarma in
Ref.~\onlinecite{scarola_dassarma_pra2005}.    
  
We define the effective exchange constant $J$ for the triple dot 
molecule with three electrons in terms of the gap between the $S=1/2$ 
and $S=3/2$ states as $E_{3/2}-E_{1/2}=3 J/ 2$.  
Without the higher order corrections $J$ would have been equal to the 
Heisenberg $J_{ij}$, otherwise it is simply related to the gap of  
the $N_e=3$ electron spectrum.   

The mapping of the behavior of our system onto the effective exchange
Hamiltonian (\ref{heishamil}) connects our analysis to the general
formalism used in quantum computing. 
\cite{loss_divincenzo_pra1998,divincenzo_bacon_nature2000}
Our considerations do not introduce any new elements into that
formalism, but rather provide means for its realistic and accurate
parametrization, reflecting the properties of an actual gated
triple-dot device. 

The Heisenberg Hamiltonian (\ref{heishamil}) can be used to model the
behavior of three electrons confined in a triple dot treated as three
coupled qubits.
However, it applies also to a coded qubit scheme, in which the
states of the entire molecule are treated as the logical states of a
single qubit.
In Ref. \onlinecite{hawrylak_korkusinski_ssc2005} we have presented a
detailed analysis of such a system, in which we selected the two
lowest total spin $1/2$ states as the logical states $|0_L\rangle$ and
$|1_L\rangle$ of the coded qubit, respectively.
In that design, the control of the energy gap between the two states
by the gate voltage provides means for the single-qubit operations. 
Again, our current work allows to parametrize this model with the
Hubbard parameters appropriate for a specific triple quantum dot.

\subsection{Comparison of Hubbard, LCAO-CI and RSP-CI results}

We shall now find the values of the Hubbard parameters appropriate for 
a typical triple quantum dot system. 
These parameters are obtained by fitting the electronic properties 
discussed above either to results of microscopic calculations, or 
to experimental data.   
In this Section we will focus on the former, while the latter will be 
discussed in Section~\ref{secexperiment}. 
 
In what follows we shall express all energies in units of the 
effective Rydberg, $1{\cal R}=m^* e^4 /2\varepsilon^2\hbar^2$, and all 
distances in units of the effective Bohr radius, 
$1a_B=\varepsilon\hbar^2/m^*e^2$,  
where $e$ and $m^*$ are the electronic charge and effective mass, 
respectively, and $\varepsilon$ is the dielectric constant of the 
material. 
For GaAs parameters, $m^*=0.067m_0$ and $\varepsilon=12.4$, we have 
$1{\cal R}=5.93$ meV and $1a_B=9.79$ nm. 
As the model lateral triple-dot system we take the structure shown in 
Fig.~\ref{fig1}(a), discussed by us in detail 
elsewhere.\cite{hawrylak_korkusinski_ssc2005}  
We take the main gray gate to be a square with the side length of 
$22.4a_B$. 
The diameter of each circular opening is $4.2a_B$, the distance 
between the centers of each pair of the holes is $4.85a_B$. 
The gate is positioned $14a_B$ above the two-dimensional electron gas 
and a voltage of $-|e|V=10{\cal R}$ is applied to it to create the 
symmetric triangular triple quantum dot. 
 
We focus on the case of $N_e=3$ confined electrons. 
Our analysis consists of two steps. 
First, we find $N_S$ lowest-lying single-particle energies and wave 
functions of the system and obtain the Coulomb matrix elements 
involving all these states. 
Second, we calculate the three-electron eigenenergies within the 
configuration-interaction (CI) approach. 
 
The one-electron properties of the system can be derived in a  
real-space approach (RSP) involving numerical diagonalization of the  
discretized single-particle 
Hamiltonian.\cite{hawrylak_korkusinski_ssc2005}  
We compute $N_S=9$ lowest-lying levels. 
The ground state is separated from the first excited state by an 
energy gap of $0.1877{\cal R}$, while the gap between the first and 
second excited states is much smaller and equal to $0.0061{\cal R}$. 
This agrees well with the Hubbard model, predicting a degeneracy of 
the two excited states. 
Also, from the average gap between the ground and excited states, 
which in the Hubbard model equals $3|t|$, we can extract the tunneling 
parameter $t_{12}=t_{13}=t_{23}=-0.0636{\cal R}$.   
 
The RSP approach, while being accurate, is computationally intensive. 
As an alternative we consider a method based on the linear combination 
of atomic orbitals (LCAO).\cite{puerto_korkusinski_prb2007}   
To this end, we approximate the numerical triple-dot lateral 
confinement, obtained as a solution of the Poisson equation, with a 
sum of three Gaussians: 
\begin{equation} 
  V(x,y) = - \sum\limits_{i=1}^3 { 
V_0^{(i)} \exp \left(- {(x-x_i)^2 + (y-y_i)^2 \over d_i^2} \right)}. 
\end{equation} 
The pairs $(x_i,y_i)$ are coordinates of the center of each dot. 
For our symmetric triple dot a good fit is obtained for 
$d_1=d_2=d_3=2.324a_B$ and  
$V_0^{(1)}=V_0^{(2)}=V_0^{(3)}=5.864{\cal R}$. 
We seek the quantum-molecular single-particle states in the form of 
linear combinations of single-dot orbitals localized on each dot. 
To simplify the calculations, we take these orbitals to be 
harmonic-oscillator (HO) wave functions of a two-dimensional parabolic 
potential, obtained by extracting the second-order component from each 
Gaussian.  
We take one $s$-type HO orbital per dot, and solve the generalized  
single-particle eigenproblem formulated in this nonorthogonal basis 
set.\cite{puerto_korkusinski_prb2007}  
As a result, we obtain $N_S=3$ quantum-molecular levels: a 
non-degenerate ground state and a doubly degenerate excited state, 
separated by a gap of $0.0354{\cal R}$. 
This structure of levels is reproduced by the Hubbard model with the 
tunneling parameter $t=-0.0118{\cal R}$. 
Note that in the LCAO case the tunneling gap is much smaller than that 
obtained in the RSP calculation. 
This is due to the restricted LCAO basis set, which underestimates the 
overlap between the single-dot orbitals. 
The agreement between the two approaches can be  improved upon inclusion of 
the $p$ and $d$  HO orbitals in the LCAO 
basis, at the expense of clarity.\cite{puerto_korkusinski_prb2007}  
 
With the single-particle energies and the Coulomb matrix elements 
calculated using the quantum-molecular orbitals, we can now proceed to 
the CI calculation of three-electron properties. 
We create all possible configurations of the three electrons with 
total $S_z=-1/2$ on $N_S$ quantum-molecular states ($N_S=9$ for the 
RSP, and $N_S=3$ for the LCAO approach), build the many-body 
Hamiltonian matrix in the basis of these configurations, and 
diagonalize it numerically.\cite{wensauer_korkusinski_ssc2004}  
The resulting spectra are shown in the 
left-hand parts of Fig.~\ref{fig3}(b) for the RSP-CI, and in 
Fig~\ref{fig3}(c) for the LCAO-CI approach. 
From the gaps separating the levels we can extract the Hubbard 
interaction parameters. 
They are: $V=0.479{\cal R}$ and $U=1.539{\cal R}$ in the RSP-CI case, 
and  $V=0.422{\cal R}$ and $U=2.557{\cal R}$ in the LCAO-CI case. 
The resulting Hubbard three-electron spectra are plotted in the 
right-hand parts of the Figures~\ref{fig3}(b) and (c). 
 
All models predict the ground state of the system to be doubly 
degenerate and to have total spin $S=1/2$.  
This is easily understood by building the lowest-energy configuration  
with triple-dot molecular orbitals: two out of three electrons are  
placed on the orbital $|M_1\rangle$ with antiparallel spins, and the  
third electron - on one of the degenerate orbitals $|M_2\rangle$ or   
$|M_3\rangle$.  
This configuration has total spin $S=1/2$, and no spin transition is  
expected upon the change of the system's topology.  
Further, in all spectra the first excited state has total spin 
$S=3/2$. 
This is the $|a_{3/2}\rangle$ state from the previous Section, 
equivalent to the spin-polarized configuration with one electron per 
dot.  
The energy gap between the low-spin and the high-spin states, 
expressed in the language of the effective Heisenberg Hamiltonian, is 
equal to $3J/2$, with $J>0$ (the ground state is antiferromagnetic). 
 
The remaining excited states, involving doubly occupied 
configurations, are visible at higher energies.   
They are separated from the low-energy, singly-occupied states by a 
gap proportional to the Hubbard onsite interaction parameter $U$. 
This parameter is larger in the LCAO-CI approach because of the 
relatively small spatial extent of the HO basis states. 
This is consistent with the underestimated tunneling gap found earlier 
in the calculation of the single-particle spectra. 
On the other hand, the Hubbard interdot interaction parameter $V$ is 
similar in both approaches. 
In the Hubbard model, the high-energy part of the spectrum is composed 
of three doubly-degenerate states. 
The degeneracy of the lowest and the highest level within this band is 
well reproduced in both microscopic models, while the middle level 
appears to be split by a smaller gap in the LCAO-CI, and a larger gap 
in the RSP-CI spectrum. 
Finally, the RSP-CI result reveals further levels, with both total 
spin $S=1/2$ and $S=3/2$. 
Their appearance is a consequence of the extended basis, containing 
$N_S=9$ single-particle molecular states, compared to $N_S=3$ states 
in LCAO-CI and Hubbard models.  
The additional states can be accounted for systematically by including 
more than one orbital per dot in the localized basis 
set.\cite{puerto_korkusinski_prb2007}  
To conclude this analysis, we find the Hubbard model to give  
qualitatively correct results but caution has to be exercised when 
making a quantitative comparison.

Up to now we have explored the case of a symmetric triangular 
triple-dot molecule. 
Let us now tune the topology of the system using the gate $GV_{13}$. 
Figure~\ref{fig4}(a) shows the three-electron spectrum obtained with  
the RSP-CI method as a function of the voltage applied to this gate, 
and the low-energy part of this spectrum is shown in the inset to this 
Figure.\cite{hawrylak_korkusinski_ssc2005}   
In the Hubbard model, this change of topology can be accounted for by 
tuning the single-particle tunneling parameter $t_{13}$. 
The corresponding spectra are shown in Fig.~\ref{fig4}(b). 
Both approaches indicate that the change of topology of the molecule 
leads to a splitting of the two degenerate $S=1/2$ levels. 
This property is a consequence of the removal of degeneracy of the 
single-particle molecular orbitals: with two electrons forming a spin 
singlet on the orbital $|M_1\rangle$, the third electron probes the 
splitting between the levels $|M_2\rangle$ and $|M_3\rangle$ 
The ability to tune the splitting by electrostatic means only suggests 
a possible use of the two low-spin states as logical states of a 
voltage-controlled coded qubit.\cite{hawrylak_korkusinski_ssc2005}  
In the language of Hubbard configurations discussed in the previous 
Section, these states can be written as 
$|0_L\rangle = \alpha_0  
{1\over\sqrt{2}}(|a\rangle-|b\rangle)+\beta_0|\Delta_0\rangle$   
and   
$|1_L\rangle = \alpha_1  
{1\over\sqrt{6}}(|a\rangle+|b\rangle-2|c\rangle)+\beta_1|\Delta_1\rangle$,  
where  $|\Delta_0\rangle$, $|\Delta_1\rangle$ are contributions of  
the doubly-occupied configurations.    
 
\section{Charging diagram of the triple dot \label{seccharging}}  
  
We can now construct the charging diagram of the triple-dot molecule.  
For any number of electrons $N_e$ (1 to 6) and any pair of gate  
voltages, or equivalently, quantum dot energies $E_i$, we can  
establish the ground-state energy $E_{GS}(N_e)$ by diagonalizing the  
Hubbard Hamiltonian.    
We use these energies to calculate the chemical potential of the  
triple quantum dot molecule $\mu(N_e)= E_{GS}(N_e+1)- E_{GS}(N_e)$.   
When $\mu(N_e)$ equals the chemical potential $\mu_L$ of the leads,  
the $N_e+1$st electron is added to the $N_e$-electron quantum-dot  
molecule.    
This establishes the total number of electrons $N_e$ in the quantum  
dot molecule and their total spin as a function of applied voltages, or  
quantum-dot energies.   
Changes in electron numbers can be detected by the Coulomb  
blockade (CB), spin blockade, or charging  
spectroscopies.\cite{pioro_abolfath_prb2005,gaudreau_studenikin_prl2006}       
The calculated stability diagram, with Hubbard parameters extracted  
from the RSP-CI calculation for three electrons, is shown in  
Fig.~\ref{fig5}.     
 
Figure~\ref{fig5}(a) shows the addition spectrum for the triple dot on  
resonance, i.e., when all dots are characterized by the same onsite 
energies, tunneling amplitudes, and Coulomb matrix elements.   
We follow the addition spectrum as   
we change the onsite energy $E$ of each quantum dot with respect to  
the chemical potential of the leads $\mu_L=0$.  
From the condition $\mu_L = E-2|t|$, the energy $E(1)$ 
corresponding to the addition of the 
first electron equals twice the hopping matrix element, $E(1)=2|t|$. 
At this energy the first Coulomb blockade peak of the triple quantum   
dot molecule should be observed.    
  
The onsite energy corresponding to the second CB peak, i.e., when the 
second electron enters the dot, equals $E(2)=-V+{8t^2\over U-V}$.   
The energy to add the second electron, or the spacing between the  
first two peaks   
$\Delta_{12}=V+2|t|-{8t^2\over U-V}$, is proportional to the direct  
Coulomb interaction $V$ between two electrons on two different dots  
and to twice the tunneling matrix element, and is reduced by  
super-exchange interaction.    
  
The third electron enters the molecule for  
$E(3)=-2V+{3J\over 2}-2|t|-{8t^2\over U-V}$   
and the spacing of the third and second CB peak equals   
$\Delta_{23}=V+2|t|-{3J\over 2}+2{8t^2\over U-V}$.   
This spacing is proportional to $V$, $2t$, and twice the  
super-exchange, but is reduced by the spin gap of the $N_e=3$ electron  
complex, equal to $3J/2$.    
The difference between the spacing of the (2,1) and (3,2) peaks,  
$\Delta_{23}-\Delta_{12}=-{3J\over 2}+3{8t^2\over U-V}$,  
directly measures the difference between the exchange in the triply  
occupied quantum dot molecule and three times the super-exchange in a  
doubly occupied quantum dot molecule.   
  
The half-filled molecule can also be probed by adding the fourth  
electron.    
This electron enters the dot for $E(4)=-U-2V-{3J\over 2}+2|t|$.    
Since the four-electron states are the first to be built by  
doubly-occupied configurations, the corresponding CB peak is spaced  
from the one for the third electron by a large on-site Coulomb energy  
$U$.    
The separation between CB peaks equals   
$\Delta_{43}=U+2{3J\over 2}-4|t|-{8t^2\over U-V}$.  
It  reflects the triplet state of two holes, and is a measure of $U$,  
$J$, and $|t|$ but not directly $V$.    
 
The expressions for peak spacings can be used to extract the Hubbard 
parameters of the system from the measured CB spectrum of the triple 
dot on resonance. 
With these parameters we can now explore the full charging diagram  
of the molecule as a function of the onsite dot energies.  
If each of the energies $E_i$ can be varied independently, the 
resulting stability diagram is three-dimensional, and  
therefore difficult to visualize. 
This is why in the proposed device, shown in Fig.~\ref{fig1}(a), 
the dots one and two are tuned by a single gate $VG_1$ while dot three 
is tuned by gate $VG_3$.    
Figure~\ref{fig5}(b) shows the corresponding cross-section of the  
stability diagram, calculated with the RSP-CI Hubbard parameters. 
The diagram shows the regions $(E_1=E_2, E_3)$ where different  
electron numbers are stable.    
The regions are denoted by $(N_1, N_2, N_3)$  where $N_i$ is  
the number of electrons (for $N_e\leq 3$) or holes (for $N_e>3$) in  
the $i$-th dot.    
For example, $(1,1,1)$ denotes the half-filled triple dot with one   
electron in each dot.   
Additionally, the regions are color-coded to indicate the total spin  
of the molecule.  
We find that the two electrons always form a spin singlet, but the  
total spin of the two-hole system can be changed from the triplet,  
which is stable close to the resonance condition, to a singlet.  
This transition can be induced by tuning the gate voltages and does  
not require the presence of a magnetic field.

\section{Comparison of theoretical and experimental charging diagrams  
  \label{secexperiment}}   

We now turn to the comparison of theory with experiment.  
The addition diagram of a triple-dot lateral device, measured recently 
by our group,\cite{gaudreau_studenikin_prl2006} is shown in 
Fig.~\ref{fig6}(a).    
The layout of the metallic gates composing the device and the 
resulting potential minima are shown in the inset to this Figure.
A similar arrangement of five bottom gates and one top gate has been
used to define electrostatically the lateral double quantum
dot, with the quantum point contact (QPC) used as a charge detector.\cite{pioro_abolfath_prb2005,koppens_folk_science2005,petta_johnson_science2005}
To this end, a sufficiently large negative voltage is applied to the
gates 1B and 5B, as well as to the top gate (T) and the middle bottom
gate (the gate 3B).
The smaller bottom gates, 2B and 4B, are then used as plungers, i.e.,
tuned with a smaller voltage to influence each of the dots locally.
However, if a large negative voltage is applied to the small gates 2B
and 4B, and a smaller voltage - to the middle gate 3B, a structure of
{\em three} potential minima is created: the dot 1 close to the
gate 1B, the dot 3 - near the gate 5B, and the dot 2 - in the middle,
between the gates T and 3B.
In this arrangement, the dots form a linear chain, so the electrons
cannot tunnel from the surrounding 2DEG directly to the middle dot.
In the experimental addition spectrum, however, three sets of lines
with distinct slopes are detected, indicating that each dot is
connected to the leads independently.
Moreover, two sets of lines show a stronger dependence on the voltage
$V_{1B}$ than on $V_{5B}$, while one set exhibits an opposite
tendency.
This suggests a formation of a ring-type arrangement, consisting of
two dots contained on the left-hand side, and one - on the right-hand
side of the device, as shown schematically in the inset to
Fig.~\ref{fig6}(a).  
The double potential minimum on the left-hand side is created most
likely by a mesoscopic fluctuation of the background potential of the
sample.
This makes it difficult to control the dots 1 and 2 independently.
On the other hand, the proposed sample layout, shown in
Fig.~\ref{fig1}(a), results in the formation of electrostatically
defined triangular triple-dot confinement.
The design has been adapted to approximate the functionality
of the experimental device, but it can be modified to allow for
independent control of both onsite and tunneling energies, however at
the expense of a more difficult to fabricate, vertical structure of
multiple gates. 
This is why we do not suggest this sample layout as a practical gating
scheme, but use it to demonstrate the degree of control we aim to
achieve in our future designs of triple-dot systems.

In our experiment, the energy landscape of the lateral confinement is
tuned by all six gates, but the addition diagram is measured only as a
function of two gate voltages, $V_{1B}$ and $V_{5B}$.
When these voltages are set to large negative values, the system is
completely depleted of electrons.   
This corresponds to the bottom left-hand region $(0,0,0)$ of the 
stability diagram.  
In the Hubbard model this region would correspond to the onsite dot  
energies $E_i$ being larger than the chemical potential of the leads,  
i.e., the upper right-hand corner $(0,0,0)$ of the diagram in  
Fig.~\ref{fig5}(b).  
As the gate voltages are made less negative, the energies $E_i$ are  
lowered, and subsequent electrons enter the molecule.  
These addition events are detected in the QPC current $I_{QPC}$,  
reacting to the changes in the charge distribution in the system.  
The dark lines in Fig.~\ref{fig6}(a) denote the boundaries between  
regions corresponding to different stable electron numbers.  
Let us focus on the addition line composed of sections A, B,  
and C, which marks the addition of the first electron to dot 1, 2, or  
3, respectively.   
The quantum molecular character of the system is revealed by the  
curvature of this line close to the regions denoted as D and E, where  
the dots 1 and 2, or 2 and 3 are on resonance, respectively.  
It is clear that the dots 1 and 2 are coupled much more strongly than  
the dots 2 and 3.  
The dashed lines drawn in these regions connect the points of  
inflection of the addition lines of the first and second electron.  
In the absence of quantum tunneling, these lines would correspond to  
the conditions $E_1=E_2$ and $E_2=E_3$, respectively.  
Away from these regions the dots are far from resonance, and the  
electrons are added to orbitals well-localized on individual dots.  
Therefore, the asymptotes drawn with respect to the sections A, B, and  
C of the one-electron line will define the respective single-dot  
properties.   
  
As a first approximation the onsite energies $E_i$ are expressed as 
linear functions of the two gate voltages:  
\begin{equation}  
  E_i = \alpha_i V_{1B} + \beta_i V_{5B} + \gamma_i.  
  \label{onsite_to_voltages} 
\end{equation}  
Let us first focus on establishing the coefficients $\alpha_i$ and 
$\beta_i$.   
In general, we seek six coefficients, but have only five equations at  
our disposal (the asymptotes to sections A, B, and C, and the two  
resonance conditions D and E), so at least one parameter has to be  
established independently.  
In this case, however, from independent measurements we know {\em  
  three} coefficients: $\alpha_1 = \alpha_2 = -100$ meV/V, and  
$\beta_3=-100$ meV/V.  
The equality of the coefficients $\alpha_1$ and $\alpha_2$ is  
reflected by the vertical character of the line D.  
The remaining three coefficients can now be easily extracted from the  
asymptotes to the sections A, B, and C, and are  
$\beta_1=-19.0$ meV/V, $\beta_2=-26.923$ meV/V, and $\alpha_3=-22.923$
meV/V.   
  
We are now in a position to convert the charging diagram from the  
coordinates $(V_{1B},V_{5B})$ to the energy coordinates, assuming for  
the moment that the coefficients $\gamma_i$ are zero.  
Since the Hubbard parameters are of one-dot and two-dot character  
only, they can be systematically fitted by extracting the features  
involving the dot pair $(i,j)$ from the charging diagram and replotting  
them as a function of the energies $(E_i,E_j)$.  
As an example we discuss the case of $i=1,j=2$, with the translated  
charging diagram shown in Fig.~\ref{fig6}(b).  
In this Figure, the solid black lines show the experimental data,  
while the dotted horizontal and vertical lines are the asymptotes.  
As already discussed, in the upper right-hand region, corresponding to  
large values of $E_1$ and $E_2$, the system is empty.  
Starting in this region, we can decrease the energy $E_1$ while  
maintaining $E_2$ constant: this corresponds to moving horizontally  
across the diagram.  
Along the way we shall first cross the rightmost vertical asymptote,  
which will mark the addition of the first electron to dot 1 in the  
zero-coupling regime (i.e., $t_{12}=0$), thereby driving the system  
into the configuration $(1,0,0)$.  
Because of our assumption of the chemical potential of the leads  
$\mu_L=0$, the energy $E_1$, which this asymptote defines, is equal  
simply to $-\gamma_1$.  
In the similar fashion, from the region $(0,0,0)$ we can move  
vertically downwards, decreasing $E_2$ while keeping $E_1$ constant.  
Crossing of the top horizontal asymptote marks the addition of the  
first electron into the second dot, i.e., formation of a configuration  
$(0,1,0)$, and defines the parameter $\gamma_2$.  
The two asymptotes cross at a right angle, which would be an expected   
behavior of the addition lines at zero coupling.  
However, the experimental data trace a hyperbola, whose curvature is a  
direct measure of the single-particle tunneling element $t_{12}$.  
  
Let us now position ourselves in the region in which the first  
electron has entered the dot 1 (the region $(1,0,0)$, the top  
part of the diagram).  
As we move vertically downwards, we encounter the top horizontal  
asymptote.  
This line would mark the addition of the second electron, and its  
placement on the second dot, but only in the case the electrons were  
not interacting.   
Since it is necessary to compensate for the Coulomb off-site charging  
energy, the actual addition takes place at lower energy $E_2$, i.e.,  
upon crossing of the horizontal asymptote second from the top.  
The energy distance between the two horizontal asymptotes corresponds  
directly to the Hubbard parameter $V_{12}$.  
An identical value is obtained by performing an analogous analysis  
starting in the region $(0,1,0)$ (the right-hand edge of the diagram),  
and moving horizontally to the left.  
Finally, we can find the onsite Coulomb terms $U_1$ and $U_2$ by  
examining the energy differences between asymptotes marking the  
addition of the second and the third electrons.  
These terms define the size of the stability region $(1,1,0)$, as  
shown in Fig.~\ref{fig6}(b).  
By using a similar analysis for the features involving the second and  
third dots, and then the first and third dots, we can systematically  
extract all Hubbard parameters.  
In our case they are (in meV):  
$\gamma_1=-34.238$, $\gamma_2=-37.169$, $\gamma_3=-36.246$,   
$t_{12}=-0.053$, $t_{13}=t_{23}=-0.0077$,  
$V_{12}=0.4623$, $V_{13}=0.0448$, $V_{23}=0.0962$,  
$U_1=2.238$, $U_2=2.1262$, and $U_3=1.8923$.   
Figure~\ref{fig7}(a) shows the charging diagram computed with the  
Hubbard model with the above parameters as a function of the gate  
voltages.  
It coincides exactly with the experimental diagram shown in  
Fig.~\ref{fig6}(a).  
 
Note that with the dependence of the onsite energies $E_i$ on gate 
voltages defined in Eq. (\ref{onsite_to_voltages}), the triple-dot 
molecule is on resonance only for $(V_1,V_5)=(-0.363V,-0.37V)$.  
In Fig.~\ref{fig6}(a) this point is found in the $(0,0,0)$ region, and 
this is why the charging diagram is essentially a superposition of two 
double-dot diagrams, and no features unique to the resonant triple-dot 
molecule are visible. 
It has been demonstrated\cite{gaudreau_studenikin_prl2006} that by 
retuning the gates making up the device the point of resonance 
can be shifted to the region of the diagram where the electrons start 
populating the system. 
This results in the appearance of the quadruple points, in which four 
different electronic configurations are on resonance, and charge 
redistribution effects similar to those in quantum cellular automata. 
 
Now let us assume that we can control the three quantum dot energies   
in our device independently, with the dots one and two on 
resonance.  
Using the Hubbard parameters found for our device we   
compute the charging diagram as a function of $E_1=E_2, E_3$, and   
shown it in Fig.~\ref{fig7}(b).  
The computed charging diagram is similar to that in 
Fig.~\ref{fig5}(b), in which we resolve the spins of the electronic 
states.   
The results agree with our theoretical predictions, including  
the existence of the triplet four-electron phase.  
We find this phase stable across only a very small range of onsite  
energies.  
Detecting the electrostatically driven triplet-singlet transition will 
be investigated in the future.

\section{Conclusions\label{summary}}  
  
In conclusion, we presented a theory of electronic properties  
of a triple quantum dot molecule.  
The electronic properties can be understood in terms of a topological 
Hunds rule, which determines the spin of the molecule as a function of 
the filling of the electronic shell.   
  
When the three dots form a symmetric triangular molecule on resonance, 
the ground state for two electrons is a spin singlet,  
for three electrons (half-filled shell) it is an anti-ferromagnetic 
$S=1/2$ configuration, and for two holes it is a triplet. 
The topology and statistics enter through the dependence of the 
energies of states on total spin.  
For example, the singlet-triplet splitting is found to depend on 
tunneling and not on charge.   
The energetics and the charging diagram are mapped out, compared with 
experiment and analyzed in detail.  
  
We have also demonstrated that the Hubbard model is capable of  
reproducing the experimental addition spectra in a quantitative  
manner.  
We have described a systematic procedure of extracting the Hubbard  
parameters from the elements of the measured charging diagram.  
Since in the experiment the single-particle orbital energies are  
controlled by gate voltages, it should be possible to induce the  
triplet-singlet transition for a four-electron molecule purely by  
electrostatic means.  
Our preliminary calculations indicate that such a transition should be  
possible in the case of the experimental lateral triple-dot device  
used by Gaudreau {\em et al.}\cite{gaudreau_studenikin_prl2006}    
  
Future work, including requirements for new device layout, is outlined.

\newpage  
  
\begin{figure}[h]  
\caption{ (Color online) (a) Cross-sectional view of a model of the
  three coupled  gated lateral quantum dots.    
  The grey rectangular gate contains three circular openings, which  
  translate into minima of the electrostatic potential at the level of  
  the two-dimensional electron gas.   
  The red and green gates can be used to shift the potential minima   
  of the dots underneath them with respect to the rest of the system.   
  The blue gate is used to tune the tunneling barrier between dots 1   
  and 3.    
  (b) Schematic representation of the triple dot structure. }  
\label{fig1}  
\end{figure}  
  
\begin{figure}[h]  
\caption{(Color online)
  The three triplet configurations in a two-electron triple dot  
  molecule (a) and examples of singly and doubly occupied singlet   
  configurations for two electrons (b) and two holes (c). }  
\label{fig2}  
\end{figure}   
  
\begin{figure}[h]  
\caption{ (Color online)
  (a) Configurations of the three-electron quantum dot with one 
  spin-up electron occupying dot 1.   
  (b) Three-electron energy levels calculated using the RSP-CI 
  approach to the device shown in Fig.~\ref{fig1} (left-hand part) and 
  using the appropriately fitted Hubbard model (right-hand part). 
  (c) Similar spectra obtained with the LCAO-CI method} 
\label{fig3}  
\end{figure}  
  
\begin{figure}[h]  
\caption{ (Color online)
  (a) Energy levels  calculated by RSP-CI technique,  
  measured from the ground state, as a function of the voltage applied  
  to the control gate $VG_{13}$.   
  Black lines show energies of total-spin-1/2 states, the red line  
  shows the energy of the spin-3/2 state.   
  Inset shows the three lowest energies as a function of the gate  
  voltage.   
  (b) Energies of three electrons localized on three Hubbard sites as  
  a function of the tunneling amplitude $t_{13}$ measured from the 
  ground state.   
  Hubbard model parameters were extracted from exact diagonalization  
  results.}  
\label{fig4}  
\end{figure}

\begin{figure}[h]  
\caption{ (Color online)
  (a) Charging diagram of the triple quantum dot on resonance  
  as a function of energy level $E$ of each dot from $N_e=0$ to   
  $N_e=6$ electrons.   
  (b) Stability diagram $(E_1-\mu=E_2-\mu;E_3-\mu)$  of the  
  triple dot molecules with dots 1 and 2 tuned by a common gate.   
  $(N_1,N_2,N_3)$ denotes average electron occupation and   
  $(-N_1,-N_2,-N_3)$ denotes average hole occupation of each dot.   
  $\mu$ denotes the chemical potential of the leads.  
  The light gray, yellow, and brown colors mark the stability regions  
  of phases with total spin $0$, $1/2$, and $1$, respectively.}  
\label{fig5}  
\end{figure}

\begin{figure}[h]  
\caption{(a) (Color online)
  Addition spectrum of a lateral triple quantum dot  
  molecule measured by Gaudreau {\em et al.}\cite{gaudreau_studenikin_prl2006}  
  Inset shows the layout of the gates defining the triple dot.  
  (b) Elements of the triple-dot addition spectrum involving dots 1  
  and 2 only, drawn as a function of onsite energies of the two dots.  
  The Hubbard parameters can be extracted directly from this diagram  
  (see text for details).  
  }  
\label{fig6}  
\end{figure}

\begin{figure}[h]  
\caption{  
  (a) Addition spectrum of the lateral triple-dot device calculated  
  within the Hubbard model after fitting to the experimental spectra  
  shown in Fig.~\ref{fig6}(a).  
  (b) (Color online)
  The same spectrum shown in the form of a charging diagram as a  
  function of the single-dot onsite energies.  
  Light gray, yellow, and brown color marks the stability region of  
  molecules with total spin $0$, $1/2$, and $1$, respectively.  
  }  
\label{fig7}  
\end{figure}

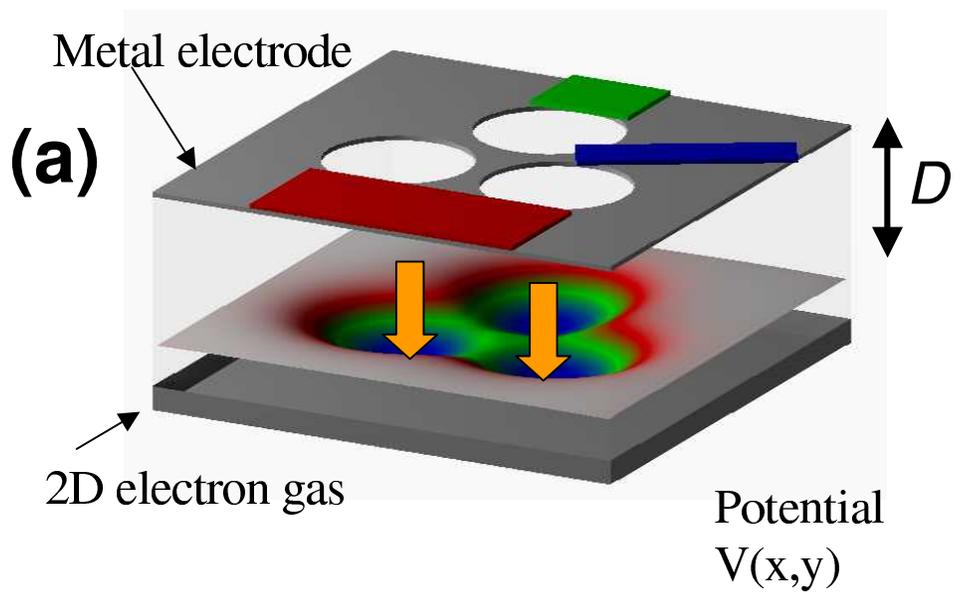
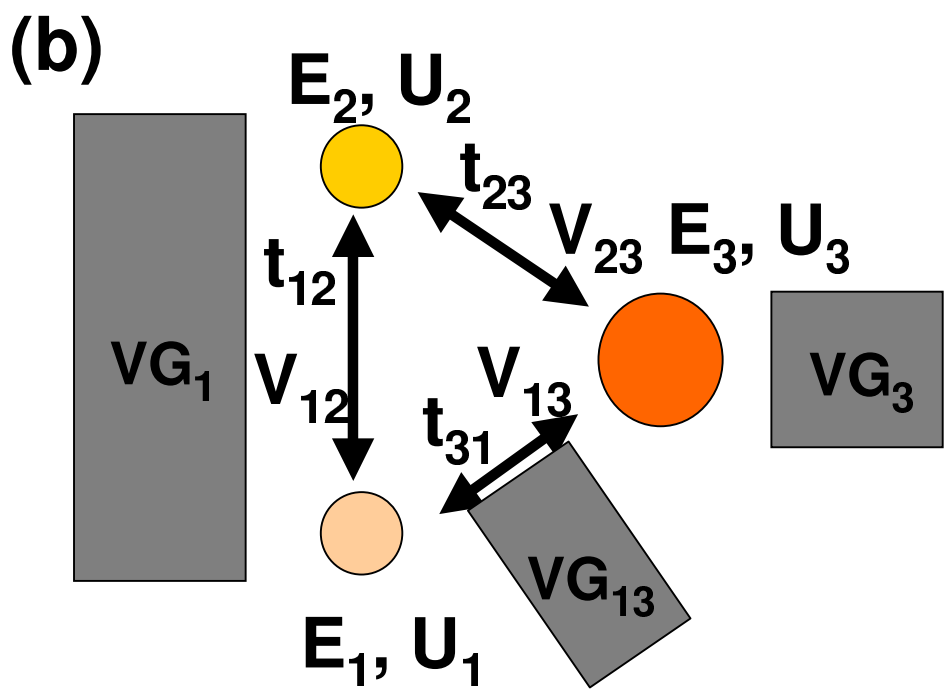
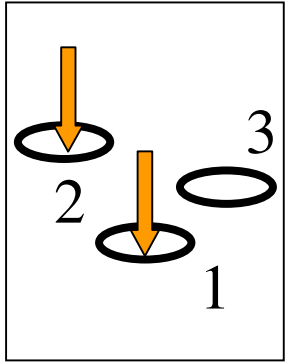
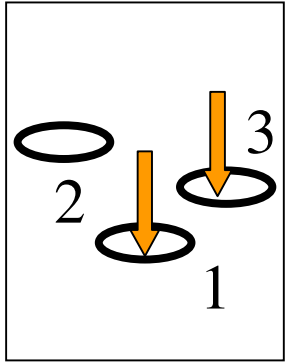
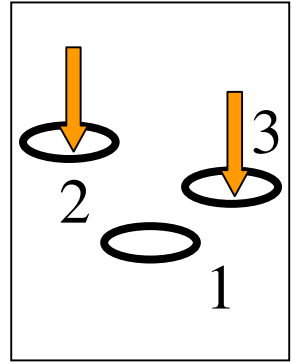
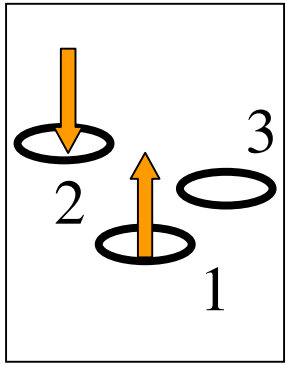
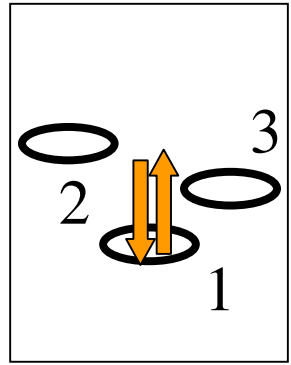
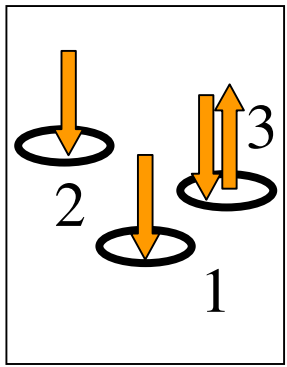
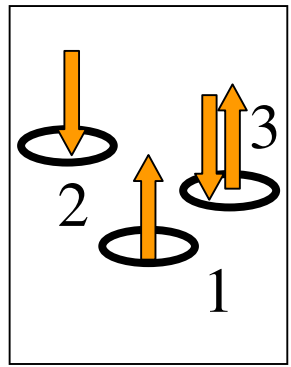
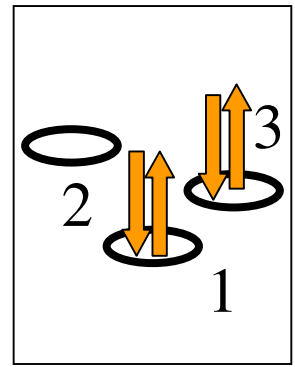
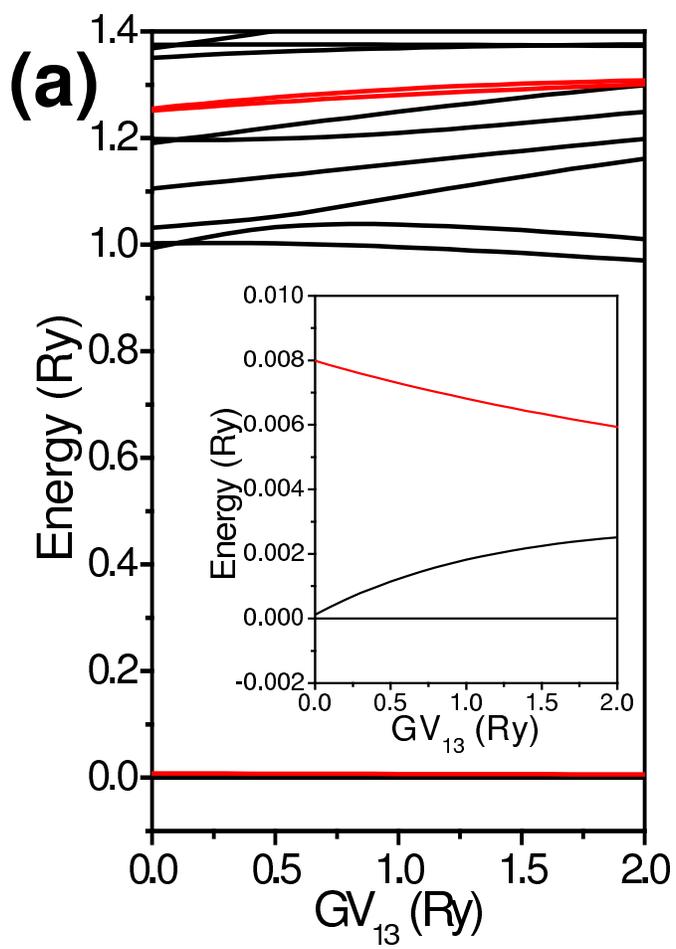
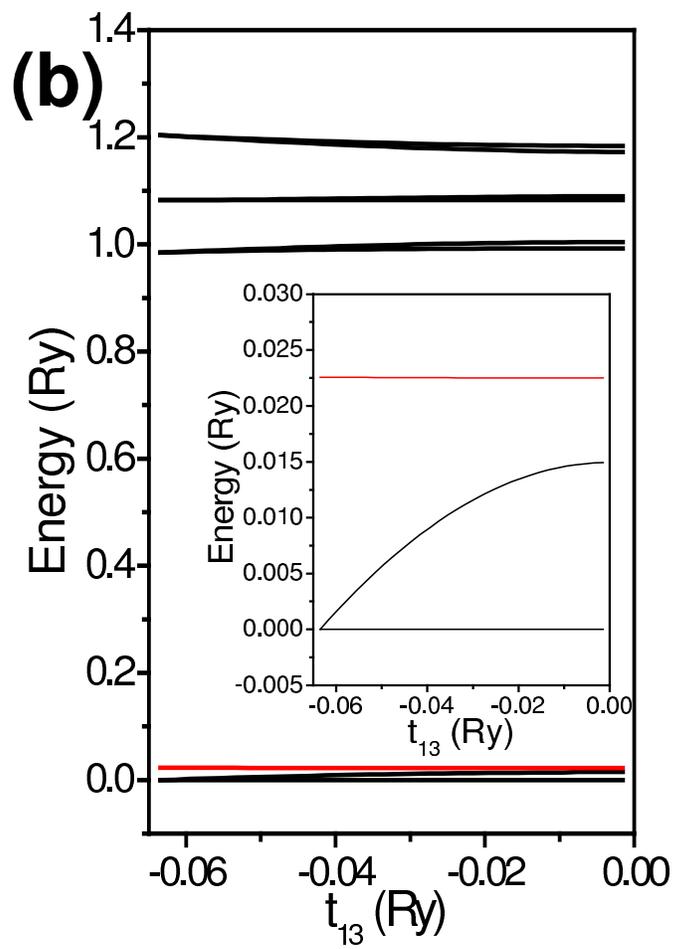
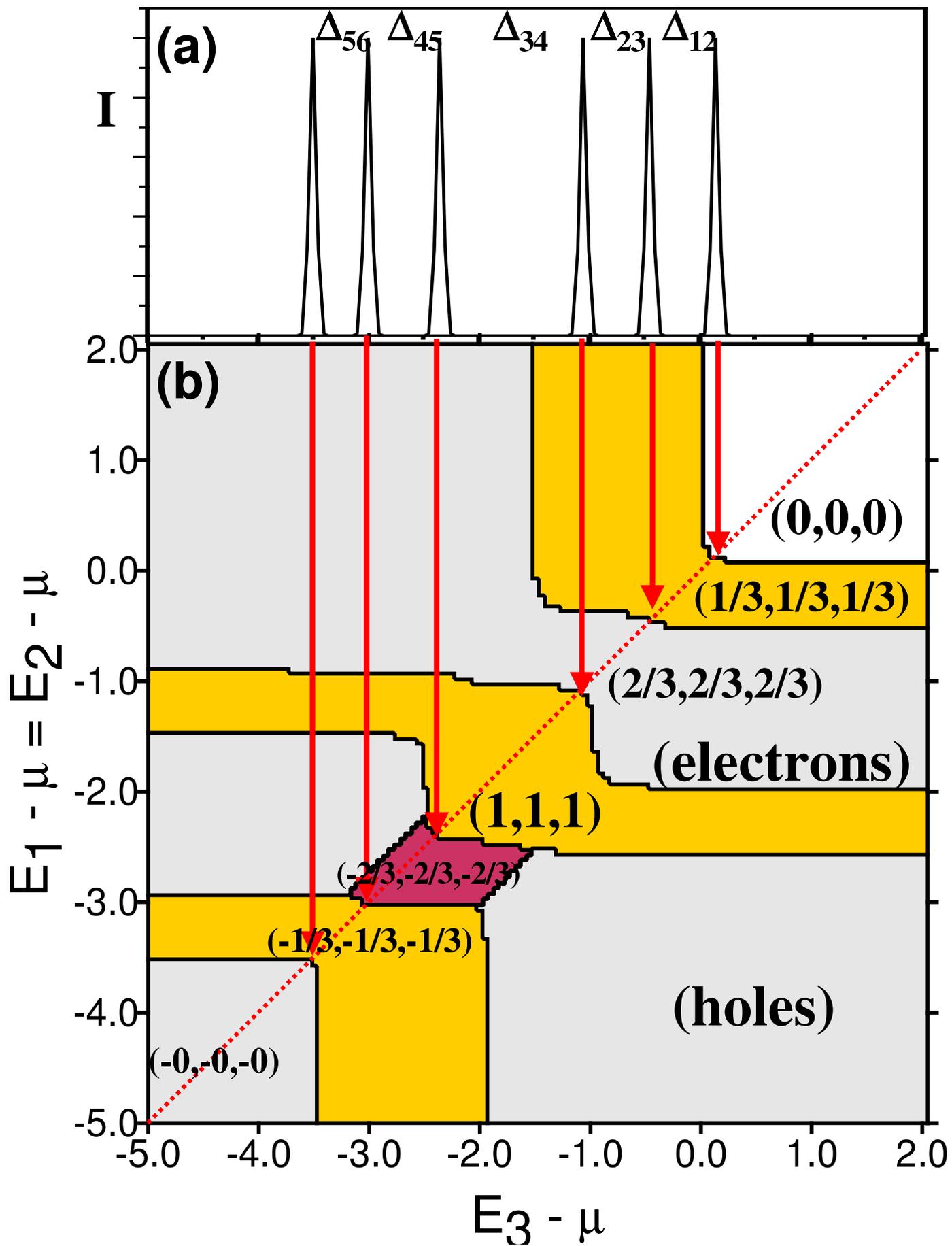
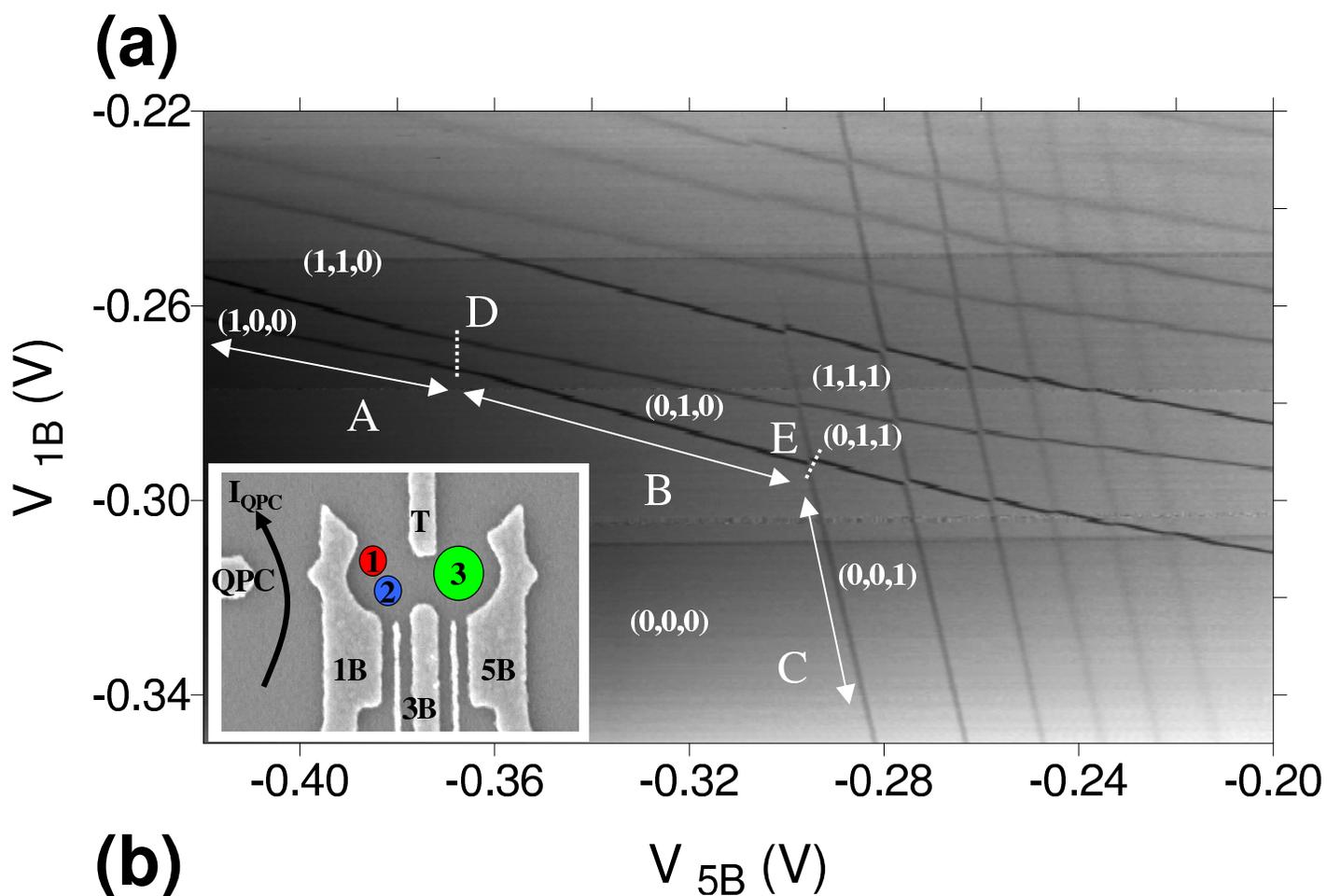
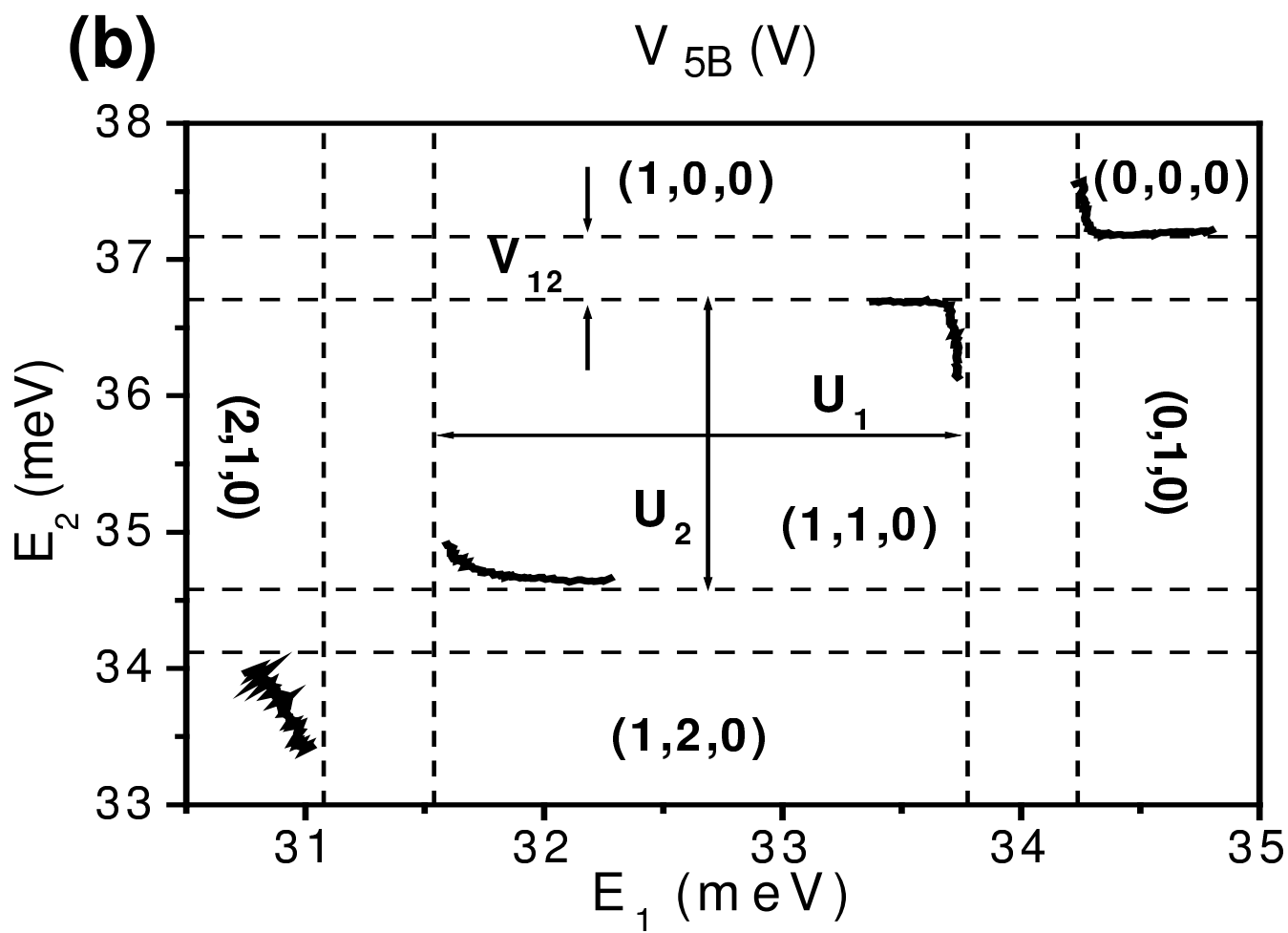
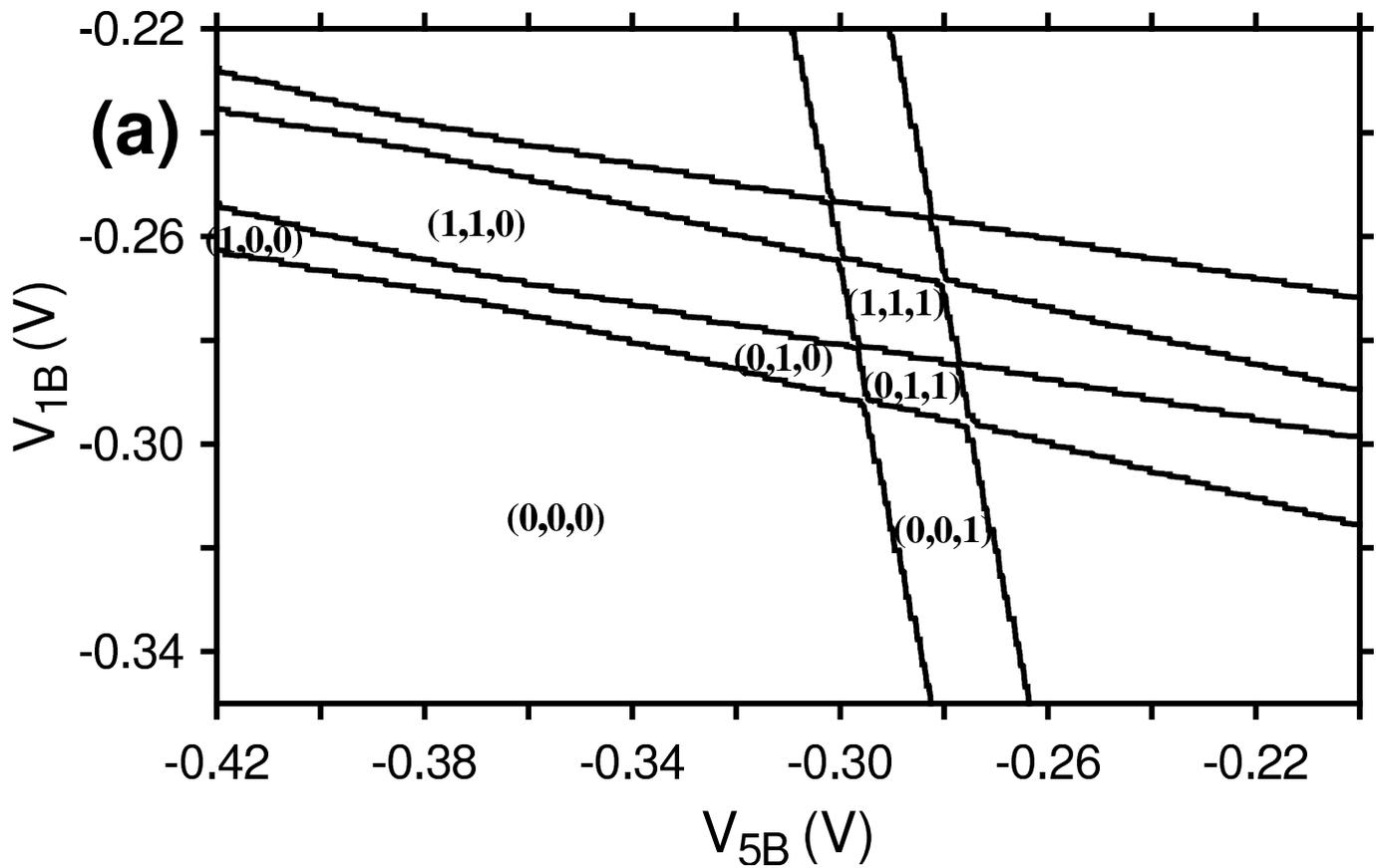
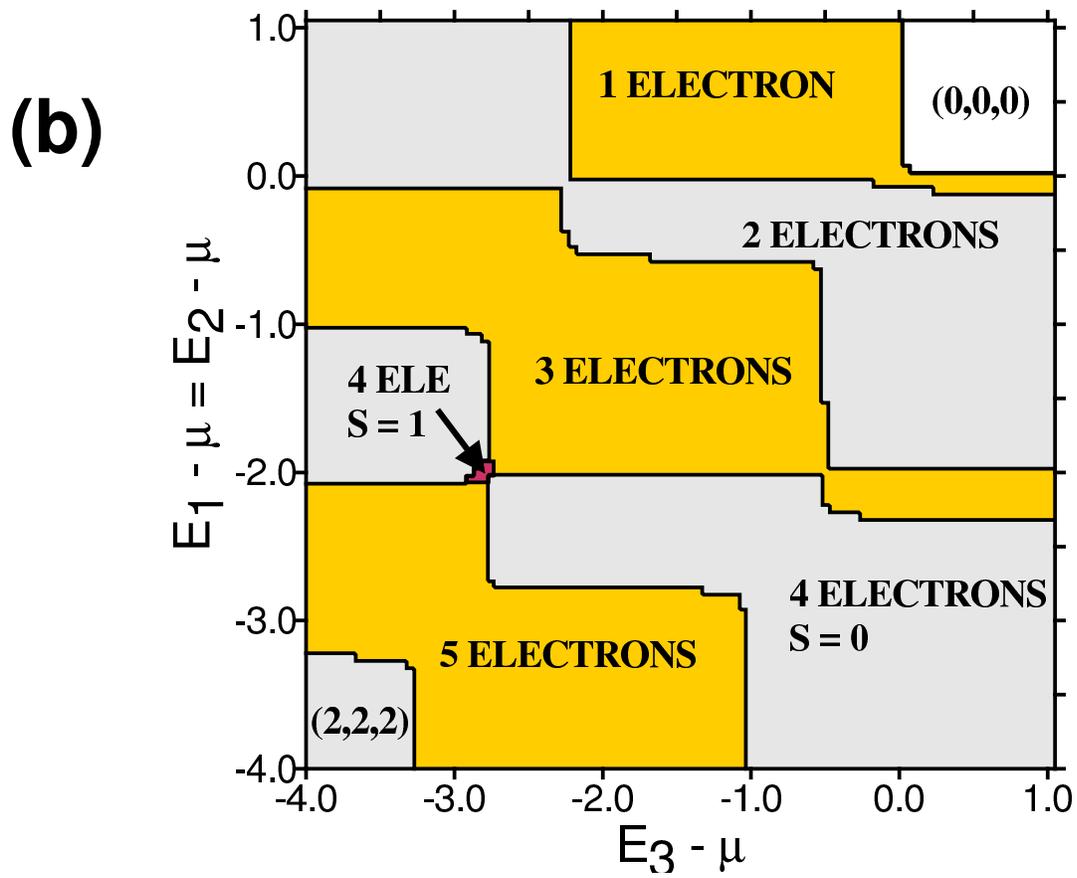
\end{document}